\documentclass[aps,final,twocolumn,showpacs,superscriptaddress]{revtex4-1}%
\usepackage{amsfonts}
\usepackage{amsmath}
\usepackage{amssymb}
\usepackage{graphicx}%
\providecommand{\U}[1]{\protect\rule{.1in}{.1in}}

\begin{document}
\title[Electron-acoustic solitary waves in the presence of a suprathermal electron component]{ Electron-acoustic solitary waves in the presence of a suprathermal electron component}
\author{Ashkbiz Danehkar}
\email{adanehkar01@qub.ac.uk}
\thanks{Present address: Department of Physics \& Astronomy, Macquarie University,
Sydney, NSW 2109, Australia}
\thanks{E-mail address:
\href{mailto:ashkbiz.danehkar@mq.edu.au}{ashkbiz.danehkar@mq.edu.au}.}
\affiliation{Centre for Plasma Physics, Department of Physics \& Astronomy, Queen's
University Belfast, Belfast BT7 1NN, Northern Ireland, United Kingdom}
\author{Nareshpal Singh Saini}
\email{ns.saini@qub.ac.uk}
\thanks{Present address: Department of Physics, Guru Nanak Dev University,
Amritsar-143005, India}
\affiliation{Centre for Plasma Physics, Department of Physics \& Astronomy, Queen's
University Belfast, Belfast BT7 1NN, Northern Ireland, United Kingdom}
\author{Manfred A. Hellberg}
\email{hellberg@ukzn.ac.za}
\affiliation{School of Physics, University of KwaZulu-Natal, Private Bag X54001, Durban
4000, South Africa}
\author{Ioannis Kourakis}
\email{i.kourakis@qub.ac.uk} \affiliation{Centre for Plasma Physics,
Department of Physics \& Astronomy, Queen's University Belfast,
Belfast BT7 1NN, Northern Ireland, United Kingdom} 

\begin{abstract}
The nonlinear dynamics of electron-acoustic localized structures in a
collisionless and unmagnetized plasma consisting of ``cool'' inertial
electrons,
``hot'' electrons having a kappa distribution, and stationary ions is
studied.
The inertialess hot electron distribution thus has a long-tailed suprathermal
(non-Maxwellian) form.
A dispersion relation is derived for linear electron-acoustic waves.
They show a strong dependence of the charge screening mechanism on excess
suprathermality (through $\kappa$). A nonlinear pseudopotential technique is
employed to investigate the occurrence of stationary-profile solitary waves,
focusing on how their characteristics depend on the spectral index $\kappa$,
and the hot-to-cool electron temperature and density ratios. Only negative
polarity solitary waves are found to exist, in a parameter region which
becomes narrower as deviation from the Maxwellian (suprathermality) increases,
while the soliton amplitude at fixed soliton speed increases. However, for a
constant value of the true Mach number, the amplitude decreases for decreasing
$\kappa$.

\end{abstract}
\volumeyear{year} \volumenumber{number} \issuenumber{number}
\eid{identifier}

\received[Received: ]{February 2, 2011}

\accepted[Accepted: ]{June 10, 2011}

\published[Published: ]{July 21, 2011}

\maketitle

\section{Introduction \label{introduction}}

Electron-acoustic waves may occur\ in plasmas characterized by a co-existence
of two distinct electron populations, here referred to as \textquotedblleft
cool\textquotedblright\ and \textquotedblleft hot\textquotedblright%
\ electrons. These are electrostatic waves of high frequency (in comparison
with the ion plasma frequency), propagating at a phase speed which lies
between the hot and cool electron thermal velocities. On such a fast (high
frequency) dynamical scale, the positive ions may safely be assumed to form a
uniform stationary charge background simply providing charge neutrality, yet
playing no essential role in the dynamics. The cool electrons provide the
inertia necessary to maintain the electrostatic oscillations, while the
restoring force comes from the hot electron pressure.

A matter of importance in electrostatic wave propagation (although inevitably
overlooked in fluid plasma models) is Landau damping, which becomes stronger
when the phase velocity approaches the thermal velocity of either electron
component, thus the wave can propagate in the plasma only within a restricted
range of parameter values. It turns out that electron-acoustic waves are
weakly damped for a temperature ratio $T_{c}/T_{h} \lesssim0.1$ and provided
that the cool electrons represent an intermediate fraction of the total
electron density: $0.2\lesssim n_{c}/(n_{c}+n_{h})\lesssim0.8$%
.~\cite{Tokar1984,Gary1985,Mace1990,Berthomier1999} The wavenumber $k$ to
minimize damping lies roughly between $0.2\lambda_{Dc}^{-1}$ and
$0.6\lambda_{Dc}^{-1}$ (where $\lambda_{Dc}$ is the cool electron Debye
length). These results on bi-Maxwellian plasmas were
later extended to include the effect of the excess suprathermality of the hot
electrons \cite{Mace1999} (a physical feature to be discussed below). It was
found that excess suprathermal electrons do cause a modification of the
damping curves, but
the overall qualitative conclusion remains unchanged: electron-acoustic waves
survive Landau damping over a wide range of parameter values.~\cite{Mace1999}
However, care must be taken in the choice of plasma configuration when
studying nonlinear electron-acoustic structures, so as to ensure that one is
considering a region of parameter space in which Landau damping is minimized.

Electron-acoustic waves occur in laboratory experiments
\cite{Henry1972,Ikezawa1981} and space plasmas, e.g., in the Earth's bow shock
\cite{Thomsen1983,Feldman1983,Bale1998} and in the auroral magnetosphere
\cite{Tokar1984,Lin1984}. They are associated with Broadband Electrostatic
Noise (BEN), a common high-frequency background activity, regularly observed
by satellite missions in the plasma sheet boundary layer (PSBL)
\cite{Matsumoto1994,Cattell1999,Kakad2009}. BEN emission includes a series of
isolated bipolar pulses, within a frequency range from $\sim10$ Hz up to the
local electron plasma frequency ($\sim10$ kHz) \cite{Matsumoto1994}. This
clearly suggests that BEN is related to electron dynamics rather than to the
ions \cite{Matsumoto1994,Kakad2009}.

In the standard bi-Maxwellian picture, the two electron species would each be
assumed to be in a (different) thermal Maxwellian distribution, parameterized
via two distinct temperature values, $T_{c}$ and $T_{h}$, respectively
\cite{Watanabe1977, Nishihara1981, Berthomier1998}. Contrary to this picture,
space and laboratory plasmas often possess an excess population of
suprathermal electrons, a fact which is reflected in a power law distribution
at high velocity (above the electron thermal speed). This excess
suprathermality phenomenon is well modeled by a generalized Lorentzian or
$\kappa$-distribution
\cite{Summers1991,Hellberg2000,Baluku2008,Hellberg2009,Pierrard2010}. The
common form of the isotropic (three-dimensional) generalized Lorentzian or
$\kappa$-distribution function is given by
\cite{Summers1991,Baluku2008,Hellberg2009}%
\begin{equation}
f_{\kappa}(v)=n_{0}(\pi\kappa\theta^{2})^{-3/2}\frac{\Gamma(\kappa+1)}%
{\Gamma(\kappa-\frac{1}{2})}\left(  1+\frac{v^{2}}{\kappa\theta^{2}}\right)
^{-\kappa-1}, \label{eq1_60}%
\end{equation}
where $n_{0}$ is the equilibrium number density of the electrons, $v$ the
velocity variable, and $\theta$ the most probable speed, which acts as a
characteristic ``modified thermal speed", and is related to the usual thermal
speed
$v_{th,e}=(2k_{B}T_{e}/m_{e})^{1/2}$ by $\theta=v_{th,e}\left[  (\kappa
-\tfrac{3}{2})/\kappa\right]  ^{1/2}$. Here $k_{B}$ is the Boltzmann constant,
$m_{e}$ the electron mass and $T_{e}$ the temperature of an equivalent
Maxwellian having the same energy content. \cite{Hellberg2009} The term
involving the Gamma function ($\Gamma$) arises from the normalization of
$f_{\kappa}(v)$, viz., $\int f_{\kappa}(v)d^{3}v=n_{0}$. Here, suprathermality
is denoted by the spectral index $\kappa$, with $\kappa> \frac{3}{2}$, for
reality of the most probable speed, $\theta$.~\cite{Hellberg2009} Low values
of $\kappa$ are associated with a significant number of suprathermal
particles; on the other hand, for $\kappa\rightarrow\infty$ a Maxwellian
distribution is recovered.

The $\kappa$-distribution was first applied to model velocity distributions
observed in space plasmas that were Maxwellian-like at lower velocities, but
had a power-law form at higher speeds
\cite{Vasyliunas1968}, and was later applied in a variety of studies,
successfully fitting many real space observations, e.g.,
\cite{Feldman1983,Christon1988,Pierrard1996,Pierrard2010}. Typical $\kappa$
values usually lie in the range $2<\kappa<6$. For example, observations in the
earth's foreshock satisfy $3<\kappa_{e}<6$, \cite{Feldman1983} measurements of
plasma sheet electron and ion distributions yield $\kappa_{i}=4.7$ and
$\kappa_{e}=$ $5.5$ (here, $e$ denotes electrons and $i$ ions),
\cite{Christon1988} and coronal electrons in the solar wind are modeled with
$2<\kappa_{e}<6$ \cite{Pierrard1996}. Recent observations of the radial
distribution of the electron population in Saturn's magnetosphere also point
towards a kappa distribution ($\kappa_{e}\simeq2.9$ - $4.2)$%
~\cite{Schippers2008}. Therefore, we focus our interest in the following on
the range $2<\kappa<6$; in fact, the Maxwellian limit is already practically
attained for values above $\kappa\simeq10$.

A linear analysis of electron-acoustic waves was first carried out
by assuming an unmagnetized Maxwellian homogeneous plasma, which
exhibited a heavily damped acoustic-like solution in addition to
Langmuir waves and ion-acoustic waves. \cite{Fried1961} Those early
results were later extended to take into account the effect of
excess suprathermal particles \cite{Mace1999,Hellberg2002}, whose
presence in fact results in an increase in the Landau damping at
small wavenumbers,
in particular when the hot electron component is dominant
\cite{Mace1999,Baluku2011}. Studies of linear and nonlinear electron-acoustic
waves in plasmas with nonthermal electrons have received a great deal of
interest in recent years
\cite{Dubouloz1991,Singh2004,Kourakis2004,Cattaert2005,Verheest2005}. Negative
potential solitary structures were shown to exist in a two-electron plasma,
either for Maxwellian \cite{Dubouloz1991} or for nonthermal
\cite{Singh2004,Sultana2010} hot electrons. Interestingly, either
incorporation of finite inertia~\cite{Cattaert2005,Verheest2005} or the
addition of a beam component~\cite{Berthomier2000,Mace2001} may lead to the
existence of positive and negative potential solitons. A recent investigation
has established the properties of modulated electron-acoustic wavepackets in
kappa-distributed plasmas, and has studied the effect of suprathermality on
the amplitude (modulational) stability.~\cite{Sultana2011}

In this paper, we study the linear and nonlinear dynamics of electron-acoustic
waves in a plasma consisting of cool adiabatic electron and hot $\kappa
$-distributed electrons, in addition to stationary ions. The paper is
organised as follows. An electron-plasma-fluid model is presented in
Section\ \ref{model}. In Section \ref{DR}, a linear dispersion relation is
derived and discussed. In Section \ref{nonlinear}, the Sagdeev pseudopotential
method is employed to investigate the occurrence of stationary profile
electrostatic solitary waves. In Section \ref{existence}, we depict the
existence domain of the electron-acoustic solitary waves. Section
\ref{investigation} is devoted to a parametric investigation of the form of
the Sagdeev pseudopotential and of the characteristics of electron-acoustic
solitary waves.
Our results are summarized in Section \ref{conclusion}.

\section{Model Equations}

\label{model}

We consider a plasma consisting of three components, namely a cool
electron-fluid (at temperature $T_{c} \neq0$), an inertialess hot
electron component with a nonthermal ($\kappa$) velocity
distribution, and uniformly distributed stationary ions.

The cool electron behavior is governed by the continuity equation,%
\begin{equation}
\frac{\partial n_{c}}{\partial t}+\frac{\partial(n_{c}u_{c})}{\partial x}=0,
\label{eq2_1}%
\end{equation}
and the momentum equation
\begin{equation}
\frac{\partial u_{c}}{\partial t}+u_{c}\frac{\partial u_{c}}{\partial x}%
=\frac{e}{m_{e}}\frac{\partial\phi}{\partial x}-\frac{1}{m_{e}n_{c}}%
\frac{\partial p_{c}}{\partial x} \, . \label{eq2_2}%
\end{equation}
The pressure of the cool electrons is governed by
\begin{equation}
\frac{\partial p_{c}}{\partial t}+u_{c}\frac{\partial p_{c}}{\partial
x}+\gamma p_{c}\frac{\partial u_{c}}{\partial x}=0 \, . \label{eq2_3}%
\end{equation}
Here $n_{c}$, $u_{c}$ and $p_{c}$ are the number density, the velocity and the
pressure of the cool electron fluid, $\phi$ is the electrostatic wave
potential, $e$ the elementary charge, and $\gamma=(f+2)/f$ denotes the
specific heat ratio (for $f$ degrees of freedom). We shall assume $\gamma=3$
(viz., $f=1$ in 1D) for the adiabatic cool electrons.

We assume the ions to be stationary (immobile), i.e., in a uniform state
$n_{i}=n_{i,0}=$ const. (where $n_{i,0}$ is the undisturbed ion density) at
all times. In order to obtain an expression for the number density of the hot
electrons, $n_{h}$, based on the $\kappa$ distribution (\ref{eq1_60}), one may
integrate Eq. (\ref{eq1_60}) over the velocity space, to obtain
\cite{Baluku2008}
\begin{equation}
n_{h}(\phi)=n_{h,0}\left(  1-\frac{e\phi}{k_{B}T_{h}(\kappa-\tfrac{3}{2}%
)}\right)  ^{-\kappa+1/2}\,, \label{eq1_1}%
\end{equation}
where $n_{h,0}$ and $T_{h}$ are the equilibrium number density and
``temperature'' of the hot electrons, respectively, and $\kappa$ is the
spectral index measuring the deviation from thermal equilibrium.

The densities of the ($\kappa$-distributed) hot electrons, the adiabatic cool
electrons, and the stationary ions are coupled via Poisson's equation:%
\begin{equation}
\frac{\partial^{2}\phi}{\partial x^{2}}=-\frac{e}{\varepsilon_{0}}\left(
Zn_{i}-n_{c}-n_{h}\right)  , \label{eq1_4}%
\end{equation}
where $\varepsilon_{0}$ is the permittivity constant, $n_{h}$ and $n_{i}$ are
the number density of hot electrons and ions, respectively.

At equilibrium, the plasma is quasi-neutral, so that
\begin{equation}
n_{c,0}+n_{h,0}=Zn_{i,0}\,,\label{eq1_5}%
\end{equation}
implying $Z{n_{i,0}}/{n_{c,0}}=1+\beta$, where we have defined the hot-to-cool
electron density ratio
\begin{equation}
\beta=\frac{n_{h,0}}{n_{c,0}}.
\end{equation}
According to Ref. \onlinecite{Tokar1984}, Landau damping is minimized in the
range $0.2\lesssim n_{c,0}/(n_{c,0}+n_{h,0})\lesssim0.8$, implying
$0.25\leqslant\beta\leqslant4$. This is our region of interest in what
follows, as nonlinear structures will not be sustainable for plasma
configurations for which the linear waves are strongly damped.

Scaling by appropriate quantities, we obtain the normalized set of equations
\begin{align}
&  \frac{\partial n}{\partial t}+\frac{\partial(nu)}{\partial x}%
=0,\label{eq2_8}\\
&  \frac{\partial u}{\partial t}+u\frac{\partial u}{\partial x}=\frac
{\partial\phi}{\partial x}-\frac{\sigma}{n}\frac{\partial p}{\partial
x},\label{eq2_9}\\
&  \frac{\partial p}{\partial t}+u\frac{\partial p}{\partial x}+3p\frac
{\partial u}{\partial x}=0,\label{eq2_10}\\
&  \frac{\partial^{2}\phi}{\partial x^{2}}=n+\beta\left[  1-\frac{\phi}%
{\kappa-3/2}\right]  ^{-\kappa+1/2}-\beta-1.\label{eq2_11}%
\end{align}
Here, $n$, $u$ and $p$ denote the cool electron fluid density, velocity and
pressure variables normalized with respect to $n_{c,0}$, $c_{th}=\left[
k_{B}T_{h}/m_{e}\right]  ^{1/2}$ and $n_{c,0}k_{B}T_{c}$, respectively. Time
and space were scaled by the plasma period $\omega_{pc}^{-1}=(n_{c,0}%
e^{2}/\varepsilon_{0}m_{e})^{-1/2}$ and the characteristic length $\lambda
_{0}=(\varepsilon_{0}k_{B}T_{h}/n_{c,0}e^{2})^{1/2}$, respectively. Finally,
$\phi$ is the wave potential scaled by $k_{B}T_{h}/e$. We have defined the
temperature ratio of the cool to the hot electrons as%
\begin{equation}
\sigma=T_{c}/T_{h}.
\end{equation}

\section{Linear waves}

\label{DR}

As a first step, we linearize Eqs.~(\ref{eq2_8})-(\ref{eq2_11}), to study
small-amplitude harmonic waves of frequency $\omega$ and wavenumber $k$. The
linear dispersion relation for electron-acoustic waves then reads:%
\begin{equation}
\omega^{2}=\frac{k^{2}}{k^{2} + k_{D, \kappa}^{2}} + 3 {\sigma} k^{2} \, ,
\label{eq2_21}%
\end{equation}
where $\sqrt{3 \sigma}$ is essentially the (normalized) cool electron thermal
velocity. After taking account of differences in normalization, this agrees
with the form found in \cite{Sahu2010}.

We note the appearance of a normalized $\kappa$-dependent screening factor
(scaled Debye wavenumber) $k_{D, \kappa}$ in the denominator, defined by
\begin{equation}
k_{D,\kappa} \equiv\dfrac{1}{\lambda_{D,\kappa}}\equiv\left[  \dfrac
{\beta(\kappa-\frac{1}{2})}{\kappa-\tfrac{3}{2}}\right]  ^{1/2} \, .
\label{eq1_22}%
\end{equation}
Since this is the inverse of the (Debye) screening length, we notice that the
latter in fact decreases due to an excess in suprathermal electrons (i.e.,
$\lambda_{D,\kappa} < \lambda_{D,\infty}$ for any finite value of $\kappa$).
This is in agreement with Refs. \onlinecite{Chateau1991,Bryant1996,Mace1998};
note also the discussion in Ref. \onlinecite{Sultana2010}.

From Eq. (\ref{eq2_21}), we see that the frequency $\omega(k)$, and hence also
the phase speed, increases with
higher temperature ratio $\sigma=T_{c}/T_{h}$. However, this is usually a
small correction to the dominant first term on the righthand side of
(\ref{eq1_22}). For large wavelength values (small $k\ll k_{D,\kappa}$), the
phase speed is given by
\begin{equation}
v_{ph}\simeq\left[  \dfrac{\kappa-\tfrac{3}{2}}{\beta(\kappa-\frac{1}{2}%
)}+3\sigma\right]  ^{1/2}\,,\label{eq2_22}%
\end{equation}
while on the other hand, the thermal contribution is dominant for high
wavenumber $k\gg k_{D,\kappa}$, i.e.%
\begin{equation}
v_{ph}\simeq(3\sigma)^{1/2}.
\end{equation}

However, we should recall from kinetic
theory\cite{Tokar1984,Gary1985,Mace1990,Mace1999} that both for very long
wavelengths ($k\lambda_{Dc}\lesssim0.2$) and very short wavelengths
($k\lambda_{Dc}\geq0.6$), the wave is strongly damped, and thus these limits
may be of academic interest only. The mode is weakly damped only for
intermediate wavelength values, where its acoustic nature is not
manifest.~\cite{Tokar1984,Gary1985,Mace1990,Mace1999,Baluku2011} Here,
$\lambda_{Dc}=(\varepsilon_{0}k_{B}T_{c}/n_{c,0}e^{2})^{1/2}$ is the cool
electron Debye length.%
\begin{figure}
[ptb]
\begin{center}
\includegraphics[
height=2.8565in,
width=2.5719in
]%
{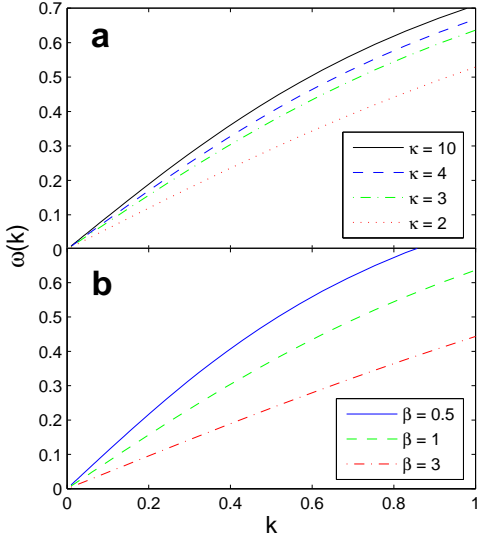}%
\caption{(Color online) Dispersion curve for harmonic (linear)
electron-acoustic waves. Upper panel: The variation of the dispersion curve
for different values of \ $\kappa$ is depicted. Curves from top to bottom:
$\kappa=10$ \ (solid), $4$\ (dashed), $3$\ (dot-dashed), and $2$\ (dotted
curve). Here, $\sigma=0.01$ and $\beta=1$. Bottom panel: Variation of the
dispersion curve for different values of \ $\beta$. Curves from top to bottom:
$\beta=0.5$\ (solid), $1$\ (dashed), and $3$\ (dot-dashed curve). Here,
$\sigma=0.01$ and $\kappa=3$.}%
\label{fig1}%
\end{center}
\end{figure}

Restoring dimensions for a moment, the dispersion relation becomes
\begin{equation}
\omega^{2}=\omega_{pc}^{2}\frac{k^{2}\lambda_{Dh}^{2}}{k^{2}\lambda_{Dh}%
^{2}+\dfrac{\kappa-\frac{1}{2}}{\kappa-\tfrac{3}{2}}}+3k^{2}c_{tc}^{2}.
\label{eq2_23}%
\end{equation}
where $c_{tc}=(k_{B}T_{c}/m_{e})^{1/2}$ is the cool electron thermal speed and
$\lambda_{Dh}$ is the hot electron Debye length defined by%
\begin{equation}
\lambda_{Dh}=\left(  \dfrac{\varepsilon_{0}k_{B}T_{h}}{n_{h,0}e^{2}}\right)
^{1/2}=\beta^{-1/2}\lambda_{0}. \label{eq1_23_1}%
\end{equation}

It appears appropriate to compare the above results with earlier results, in
the linear regime. First of all, we note that Ref. \onlinecite{Mace1999} has
adopted a kinetic description of electron-acoustic waves in suprathermal
plasmas. For this purpose, Eq.~(\ref{eq2_23}) may be cast in the form,
\begin{align}
\omega^{2}  &  =k^{2}\left(  \frac{C_{s\kappa}^{2}}{1+k^{2}\lambda_{\kappa
h}^{2}}+3c_{tc}^{2}\right) \nonumber\\
&  =\omega_{pc}^{2}\frac{1+3k^{2}\lambda_{Dc}^{2}+3\lambda_{Dc}^{2}%
/\lambda_{\kappa h}^{2}}{1+1/k^{2}\lambda_{\kappa h}^{2}}\,. \label{eq2_24}%
\end{align}
Here $C_{s\kappa}^{2}=\omega_{pc}^{2}\lambda_{\kappa h}^{2}=(n_{c,0}%
/n_{h,0})[(\kappa-3/2)/(\kappa-1/2)]c_{th}^{2}$ is the electron acoustic speed
in a suprathermal plasma, while the effective shielding length is given by
$\lambda_{\kappa h}^{2}=[\epsilon_{0}k_{B}T_{h}/n_{h,0}e^{2}][(\kappa
-3/2)/(\kappa-1/2)]$.
Our Eq. (\ref{eq2_24}) above agrees with relation (3) in Ref.
\onlinecite{Mace1999} (upon considering the limit $\lambda_{Dc}\rightarrow0$
therein). In the limit $\kappa\rightarrow\infty$ (Maxwellian distribution),
Eq.~(\ref{eq2_24}) recovers precisely
%
Eq.~(1) in Ref. \onlinecite{Berthomier2000}, upon considering the limit
$\lambda_{Dc}\rightarrow0$. Furthermore, one recovers exactly Eq.~(5) in Ref.
\onlinecite{Kakad2009} for Maxwellian plasma in the cold-electron limit
($T_{c}{=0}$).

In Figure \ref{fig1}, we depict the dispersion curve of the electron-acoustic
mode, showing the effect of varying the values of the spectral index $\kappa$
and the density ratio $\beta$. It is confirmed numerically that the phase
speed ($\omega/k$) increases weakly with a reduction in suprathermal particle
excess, as the Maxwellian is approached, and that there is a significant
reduction in phase speed as the plasma model changes from one in which the
cool electrons dominate, to one which is dominated by the hot electron density.

\section{Nonlinear analysis for large amplitude solitary waves}

\label{nonlinear}

Anticipating constant profile solutions, we shall consider Eqs. (\ref{eq2_8}%
)--(\ref{eq2_11}) in a stationary frame traveling at a constant normalized
velocity $M$ (to be referred to as the Mach number), implying the
transformation $\xi=x-Mt$. The space and time derivatives are thus replaced by
$\partial/\partial x=d/d\xi$ and $\partial/\partial t=-Md/d\xi$, respectively,
so Eqs. (\ref{eq2_8})--(\ref{eq2_11}) take the form:%
\begin{equation}
-M\dfrac{dn}{d\xi}+\frac{d(nu)}{d\xi}=0,\label{eq2_27}%
\end{equation}%
\begin{equation}
-M\dfrac{du}{d\xi}+u\dfrac{du}{d\xi}=\dfrac{d\phi}{d\xi}-\frac{\sigma}%
{n}\dfrac{dp}{d\xi},\label{eq2_28}%
\end{equation}%
\begin{equation}
-M\dfrac{dp}{d\xi}+u\dfrac{dp}{d\xi}+3p\dfrac{du}{d\xi}=0,\label{eq2_28_1}%
\end{equation}%
\begin{equation}
\frac{d^{2}\phi}{d\xi^{2}}=-(\beta+1)+n+\beta\left[  1-\frac{\phi}%
{(\kappa-\tfrac{3}{2})}\right]  ^{-\kappa+1/2}.\label{eq2_29}%
\end{equation}
We assume that the equilibrium state is reached at both infinities
($\xi\rightarrow\pm\infty$). Accordingly, we integrate and apply the boundary
conditions $n=1$, $p=1$, $u=0$ and $\phi=0$ at $\pm\infty$. One thus obtains%
\begin{equation}
u=M\left(  1-\frac{1}{n}\right)  ,\label{eq2_30}%
\end{equation}%
\begin{equation}
u={M-(M}^{2}{+2\phi-3n^{2}\sigma+3\sigma)}^{1/2},\label{eq2_30_1}%
\end{equation}
and%
\begin{equation}
p=n^{3}.\label{eq2_31}%
\end{equation}

Combining Eqs. (\ref{eq2_30})--(\ref{eq2_31}), we obtain the following
biquadratic equation for the cool electron density,%
\begin{equation}
{3\sigma n^{4}}-({M}^{2}{+2\phi+3\sigma)n^{2}}+{M}^{2}=0\,. \label{eq2_32_1}%
\end{equation}
The solution of Eq. (\ref{eq2_32_1}) may be written as
\begin{equation}
{n=}\dfrac{1}{2}\left(  n_{(+)}\pm n_{(-)}\right)  , \label{eq2_32}%
\end{equation}
where
\begin{align}
n_{(+)}  &  {\equiv}\left[  \dfrac{{2\phi+}\left(  {M+}\sqrt{3{\sigma}%
}\right)  ^{2}}{3{\sigma}}\right]  ^{1/2},\\
n_{(-)}  &  {\equiv}\left[  \dfrac{{2\phi+\left(  {M-}\sqrt{3{\sigma}}\right)
^{2}}}{3{\sigma}}\right]  ^{1/2}. \label{eq2_32_2}%
\end{align}
From the boundary conditions, $n=1$ at $\phi=0$, it follows that the negative
sign must be taken in Eq. (\ref{eq2_32}). Furthermore, we shall assume that
$M>\sqrt{3\sigma}$, i.e., that the cool electrons are supersonic, while the
hot electrons are subsonic, thus we require that $M<1$.

Reality of the density variable imposes the requirement ${2\phi+\left(
{M-}\sqrt{3{\sigma}}\right)  ^{2}>0}$, which implies a limit
on the electrostatic potential value
$|\phi_{\max}|=\frac{1}{2}\left(  {M-}\sqrt{3{\sigma}}\right)  ^{2}$
associated with negative solitary structures (positive electric potentials,
should they exist, satisfy the latter condition automatically, and are thus
not limited).

Substituting the density expression (\ref{eq2_32})--(\ref{eq2_32_2}) into
Poisson's equation (\ref{eq2_29}) and integrating,
yields the pseudo-energy balance equation for a unit mass in a conservative
force field, if one defines $\xi$ as \textquotedblleft time\textquotedblright%
\ and $\phi$ as \textquotedblleft position\textquotedblright\ variable:%
\begin{equation}
\frac{1}{2}\left(  \frac{d\phi}{d\xi}\right)  ^{2}+\Psi(\phi)=0,
\label{eq2_37}%
\end{equation}
where the Sagdeev pseudopotential $\Psi(\phi)$ is given by
\begin{align}
\Psi(\phi)  &  =\beta\left[  1-\left(  1+\frac{\phi}{-\kappa+\tfrac{3}{2}%
}\right)  ^{-\kappa+3/2}\right]  +(1+\beta)\phi\nonumber\\
&  +\frac{1}{6\sqrt{3{\sigma}}}\left[  \left(  {M+}\sqrt{3{\sigma}}\right)
^{3}-{{\left(  {M-}\sqrt{3{\sigma}}\right)  ^{3}}}\right. \nonumber\\
&  -\left(  {2\phi+}\left[  {M+}\sqrt{3{\sigma}}\right]  ^{2}\right)
^{3/2}\nonumber\\
&  \left.  +{\left(  {2\phi+\left[  {M-}\sqrt{3{\sigma}}\right]  ^{2}}\right)
}^{3/2}\right]  . \label{eq2_38}%
\end{align}%
\begin{figure}
[ptb]
\begin{center}
\includegraphics[
height=4.6406in,
width=2.4664in
]%
{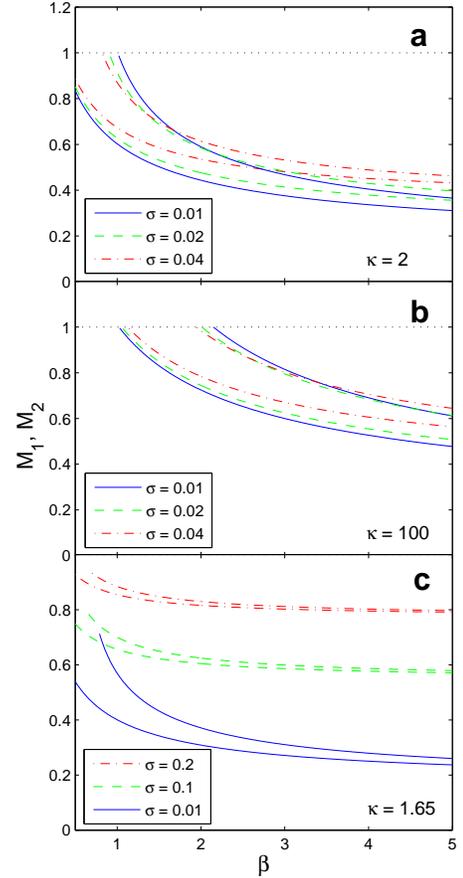}%
\caption{(Color online) Variation of the lower limit $M_{1}$ (lower curves)
and the upper limit $M_{2}$ (upper curves) with the hot-to-cold electron
density ratio $\beta$ for different values of the temperature ratio $\sigma$.
Solitons may exist for values of the Mach number $M$ in the region between the
lower and the upper curve(s) of the same style/color. Curves: (a-b)
$\sigma=0.01$\ (solid), $0.02$\ (dashed), and $0.04$\ (dot-dashed), and (c)
$\sigma=0.01$\ (solid), $0.1$\ (dashed), and $0.2$\ (dot-dashed) Here, we have
taken: (a) $\kappa=2$, (b) $\kappa=100$ (quasi-Maxwellian), and (c)
$\kappa=1.65$. }%
\label{fig2}%
\end{center}
\end{figure}
\begin{figure}
[ptb]
\begin{center}
\includegraphics[
height=3.3122in,
width=2.6792in
]%
{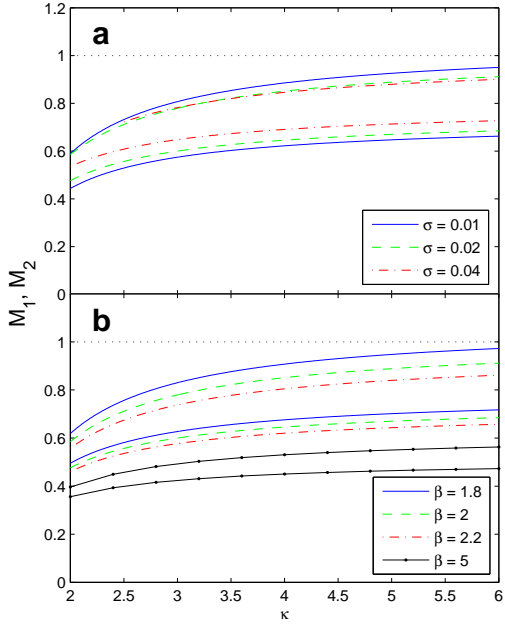}%
\caption{(Color online) Variation of the lower limit $M_{1}$ (lower curves)
and the upper limit $M_{2}$ (upper curves) with the suprathermality parameter
$\kappa$ for different values of the temperature ratio $\sigma$ (upper panel),
and density ratio $\beta$ (bottom panel). Solitons may exist for values of the
Mach number $M$ in the region between the lower and upper curves of the same
style/color. Upper panel: $\sigma=0.01$\ (solid curve), $0.02$\ (dashed), and
$0.04$\ (dot-dashed). Here, we have taken $\beta=2$. Lower panel: $\beta
=1.8$\ (solid), $2$\ (dashed), $2.2$\ (dot-dashed), and $5$\ (solid circles).
Here, $\sigma=0.02$.}%
\label{fig3}%
\end{center}
\end{figure}

\section{Soliton existence domain}

\label{existence}

We next investigate the conditions for existence of solitons. First, we need
to ensure that the origin at $\phi=0$ is a root and a local maximum of $\Psi$
in Eq. (\ref{eq2_38}), i.e., $\Psi(\phi)=0$, $\Psi^{\prime}(\phi)=0$ and
$\Psi^{\prime\prime}(\phi)<0$ at $\phi=0$
\cite{Verheest2004,McKenzie2004,Verheest2007}, where primes denote derivatives
with respect to $\phi$. It is easily seen that the first two constraints are
satisfied. We thus impose the condition
\begin{equation}
F_{1}(M)=-\left.  \Psi^{\prime\prime}(\phi)\right\vert _{\phi=0}=\frac
{\beta(\kappa-\frac{1}{2})}{\kappa-\tfrac{3}{2}}-\frac{1}{M^{2}-3{\sigma}}>0.
\label{eq2_44}%
\end{equation}
Eq. (\ref{eq2_44}) provides the minimum value for the Mach number, $M_{1}$,
i.e. :%
\begin{equation}
M>M_{1}=\left[  \frac{\kappa-\tfrac{3}{2}}{\beta(\kappa-\frac{1}{2})}%
+3{\sigma}\right]  ^{1/2}. \label{eq2_45}%
\end{equation}
Clearly, $M_{1}$ is
the (normalized) electron-acoustic phase speed -- cf. Eq.~(\ref{eq2_22}). It
is thus also related to Debye screening via the screening parameter
$\lambda_{D,\kappa}$ in (\ref{eq1_22}), associated with the hot $\kappa
$-distributed electrons.
We deduce that soliton solutions are super-acoustic.
For Maxwellian hot electrons ($\kappa\rightarrow\infty$) and cold
\textquotedblleft cool\textquotedblright\ electrons ($\sigma=0$), we obtain
$M_{1}=1/\beta^{1/2}$, thus recovering the normalized phase speed for
electron-acoustic waves in a Maxwellian plasma.~\cite{Mace1999} The lower Mach
number limit, $M_{1}$, increases with $T_{c}$ (via $\sigma$), and decreases
for lower values of $\kappa$ (large excess of suprathermal electrons), and
hence the sound speed in suprathermal plasmas is
reduced, in comparison with Maxwellian plasmas ($\kappa\rightarrow\infty$).

An upper limit for $M$ \ is found through the fact that the cool electron
density becomes complex at $\phi=\phi_{max}$, and hence the largest soliton
amplitude satisfies $F_{2}(M)=\Psi(\phi)|_{\phi=\phi_{\max}}>0$. This yields
the following equation for the upper limit in M:%
\begin{align}
F_{2}(M)  &  =-\frac{1}{2}(1+\beta)\left(  {M-}\sqrt{3{\sigma}}\right)
^{2}-\frac{4}{3}M^{3/2}\left(  3{\sigma}\right)  ^{1/4}\nonumber\\
&  +\beta\left(  1-\left[  1+\frac{\left(  {M-}\sqrt{3{\sigma}}\right)  ^{2}%
}{2\kappa-3}\right]  ^{-\kappa+3/2}\right) \nonumber\\
&  +M{{^{2}+}\sigma} =0\,. \label{eq2_47}%
\end{align}
Solving Eq. (\ref{eq2_47}) provides the upper limit $M_{2}(\kappa,\beta
,\sigma)$ for acceptable values of the Mach number for solitons to exist.

For comparison, for a Maxwellian distribution (here recovered as
$\kappa\rightarrow\infty$), the constraints reduce to
\begin{align}
F_{1}(M)  &  =\beta-\frac{1}{M^{2}-3{\sigma}}>0,\\
F_{2}(M)  &  =-\frac{1}{2}(1+\beta)\left(  {M-}\sqrt{3{\sigma}}\right)
^{2}-\frac{4}{3}M^{3/2}\left(  3{\sigma}\right)  ^{1/4}\nonumber\\
&  +\beta\left(  1-\exp\left[  -\tfrac{1}{2}\left(  {M-}\sqrt{3{\sigma}%
}\right)  ^{2}\right]  \right) \nonumber\\
&  +M{{^{2}+}\sigma}>0.
\end{align}
The latter equation provides the upper limit $M_{2}$, while the lower limit
becomes $M_{1}=(1/\beta+3\sigma)^{1/2}$.

In the opposite limit of ultrastrong suprathermality, i.e., $\kappa
\rightarrow3/2$, the Mach number threshold approaches a non-zero limit
$M_{1}=\sqrt{3{\sigma}}$, which is essentially the thermal speed, as noted
above (recall that $M>\sqrt{3{\sigma}}$ by assumption). The upper limit
$M_{2}$ is then given by
\begin{align}
F_{2}(M)  &  =-\frac{1}{2}(1+\beta)\left(  {M-}\sqrt{3{\sigma}}\right)
^{2}+M{{^{2}+}\sigma}\nonumber\\
&  -\frac{4}{3}M^{3/2}\left(  3{\sigma}\right)  ^{1/4}=0\,.
\end{align}
Interestingly, the two limits $M_{1}$\ and $M_{2}$ both tend to the same limit
as $\kappa\rightarrow3/2$,
namely, $\sqrt{3{\sigma}}$, where the soliton existence region vanishes, as
the kappa distribution breaks down.

We have studied the existence domain of electron-acoustic solitary waves for
different values of the parameters. The results are depicted in
Figs.~\ref{fig2}--\ref{fig3}.
Solitary structures of the electrostatic potential may occur in the range
$M_{1}<M<M_{2}$, which depends on the parameters $\beta$, $\kappa$, and
$\sigma$. We recall that we have also assumed that cool electrons are
supersonic (in the sense $M>\sqrt{3{\sigma}}$)
\cite{Verheest2004,McKenzie2004,Verheest2007}, and the hot electrons subsonic
($M<1$), and care must be taken not to go beyond the limits of the plasma
model.%
\begin{figure}
[ptb]
\begin{center}
\includegraphics[
height=3.4722in,
width=2.5573in
]%
{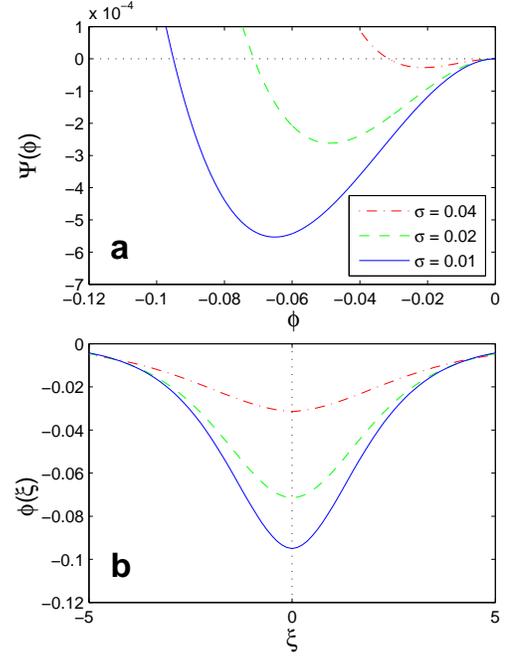}%
\caption{(Color online) The pseudopotential $\Psi(\phi)$\ (upper panel) and
the associated solution (electric potential pulse) $\phi$ (lower panel) are
depicted versus position $\xi$, for different values of the temperature ratio
$\sigma\,$. We have taken: $\sigma=0.01$ (solid curve), $0.02$ (dashed curve),
and $0.04$ (dot-dashed curve). The other parameter values are $\beta=1.3$,
$\kappa=2.5$\ and $M=0.75$.}%
\label{fig5}%
\end{center}
\end{figure}

The interval $[M_{1},M_{2}]$ where solitons may exist is depicted in Fig.
\ref{fig2}, in two opposite cases: in (a) and (c) two very low, and in (b) one
very high value of $\kappa$. We thus see that for both a quasi-Maxwellian
distribution and one with a large excess suprathermal component of hot
electrons, both $M_{1}$ and $M_{2}$ decrease with an increase in the relative
density parameter $\beta$ for fixed $\kappa$ and soliton speed $M$. Further,
the upper limit falls off more rapidly, and thus the existence domain in Mach
number becomes narrower for higher values of the hot-to-cool electron density
ratio. Comparing the two frames (a) and (b) in Fig. \ref{fig2}, we immediately
notice that suprathermality (low $\kappa$) results in solitons propagating at
lower Mach number values, a trend which is also seen in Fig.~\ref{fig2}c.
Another trend that is visible in Figs.~\ref{fig2}--\ref{fig3}a is that
increased thermal pressure effects of the cool electrons, manifested through
increasing $\sigma$, also lead to a narrowing of the Mach number range that
can support solitons. Finally, we note that for $\beta\sim1$, the upper limit
found from Eq.~(\ref{eq2_47}) rises above the limit $M=1$ required by the
assumptions of the model, and the latter then forms the upper limit.

Interestingly, in Figs. \ref{fig2}--\ref{fig3} the existence region appears to
shrink down to nil, as the curves approach each other for high $\beta$ values.
This is particularly visible in Fig. \ref{fig2}c, for a very low value of
$\kappa$ ($\kappa=1.65$). This is not an unexpected result, as high values of
$\beta$ are equivalent to a reduction in cool electron relative density, which
leads to our model breaking down if the inertial electrons vanish. We recall
that a value $\beta>4$ is a rather abstract case, as it corresponds to a
forbidden regime, since Landau damping will prevent electron-acoustic
oscillations from propagating. Similarly, a high value of the temperature
ratio, such as $\sigma=0.2$, takes us outside the physically reasonable
domain. Nevertheless, as it appears that the lower and upper limits in $M$
approach each other asymptotically for high values of $\beta$, we have carried
out calculations for increasing $\beta$, up to $\beta=100$ for $\sigma=0.2$ as
an academic exercise, and can confirm that the two limits do not actually
intersect.%
\begin{figure}
[ptb]
\begin{center}
\includegraphics[
height=5.7545in,
width=2.5097in
]%
{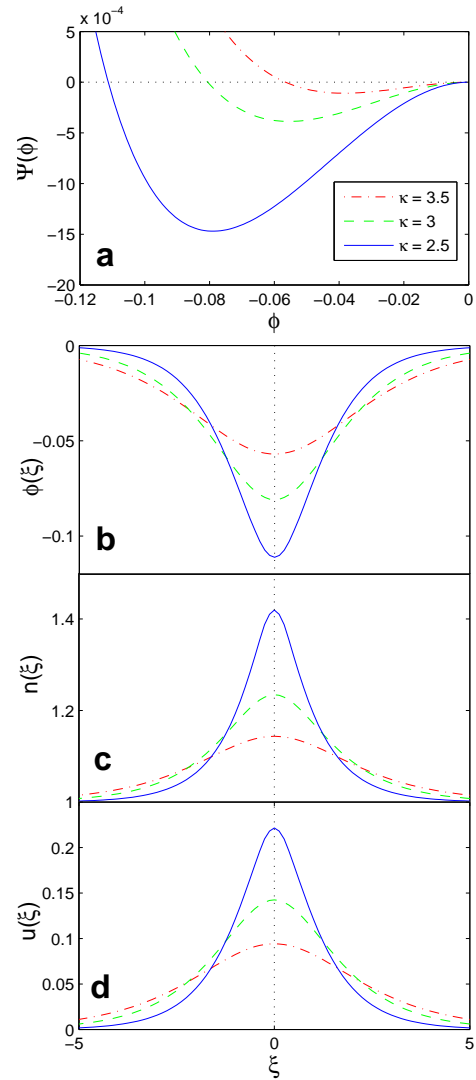}%
\caption{(Color online) (a) The pseudopotential $\Psi(\phi)$\ and the
associated solutions: (b) electric potential pulse $\phi$, (c) density $n$,
and (d) velocity $u$ are depicted versus position $\xi$, for different
$\kappa$. We have taken: $\kappa=2.5$\ (solid curve), $3$\ (dashed curve), and
$3.5$\ (dot-dashed curve). The other parameter values are: $\sigma=0.02$,
$\beta=1.6$,\ and $M=0.75$.}%
\label{fig6}%
\end{center}
\end{figure}

Figure \ref{fig3} shows the range of allowed Mach numbers as a function of
$\kappa$, for various values of the temperature ratio $\sigma$. As discussed
above, increasing $\kappa$ towards a Maxwellian distribution ($\kappa
\rightarrow\infty$) broadens the Mach number range and yields higher values of
Mach number. On the other hand, both upper and lower limits decrease as the
limiting value
$\kappa\rightarrow3/2$ is approached. The qualitative conclusion is analogous
to the trend in Fig. \ref{fig2}: stronger excess suprathermality leads to
solitons occurring in narrower ranges of $M$. Furthermore, as illustrated in
Figs. \ref{fig2} and \ref{fig3}a, the Mach number threshold $M_{1}$ approaches
the upper limit $M_{2}$ for high values of $\sigma$ and $\beta$: both
increased hot-electron density and cool-electron thermal effects shrink the
permitted soliton existence region.

Figure \ref{fig3}b depicts the range of allowed Mach numbers as a function of
$\kappa$ for various values of the density parameter $\beta$ (for a fixed
indicative $\sigma$ value). We note that both curves decrease with an increase
in $\beta$. Although it lies in the damped region, we have also depicted a
high $\beta$ regime for comparison (solid-crosses curve).

We conclude this section with a brief comparison of our work with that of Ref.
\onlinecite{Sahu2010}. The latter did not consider existence domains at all,
let alone their dependence on plasma parameters, but merely plotted some
Sagdeev potentials and associated soliton potential profiles for chosen values
of some of the parameters, so as to extract some trends. En passant, there is
indirect mention of an upper limit in $M$, in that it is commented that as
increasing values of $M$ are considered, at some stage solitary waves cease to
exist.~\cite{Sahu2010}%
\begin{figure}
[ptb]
\begin{center}
\includegraphics[
height=2.111in,
width=3.1384in
]%
{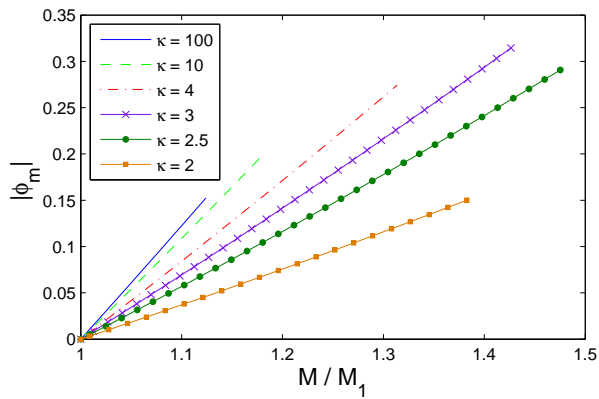}%
\caption{(Color online) The dependence of the pulse amplitude $|\phi_{m}|$ on
the Mach number-to-sound-speed ratio $M/M_{1}$ is depicted, for different
values of $\kappa$. From top to bottom: $\kappa=100$\ (solid curve);
$10$\ (dashed curve); $4$\ (dot-dashed curve); $3$\ (crosses); $2.5$\ (solid
circles); $2$\ (solid squares). Here, $\sigma=0.01$ and $\beta=1.3$.}%
\label{fig8}%
\end{center}
\end{figure}
\begin{figure}
[ptb]
\begin{center}
\includegraphics[
height=3.2534in,
width=2.5028in
]%
{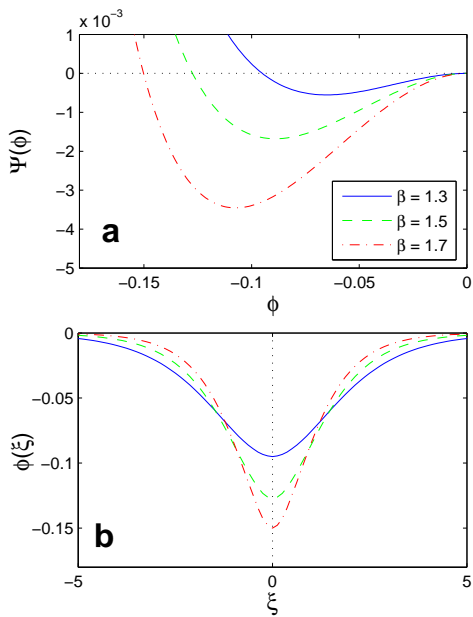}%
\caption{(Color online) (Upper panel) The pseudopotential $\Psi(\phi)$ vs.
$\phi$\ and (lower panel) the associated electric potential pulse $\phi$ vs.
$\xi$\ are depicted, for different values of the hot-to-cold electron density
ratio $\beta$. From top to bottom: $\beta=1.3$\ (solid curve); $1.5$ (dashed
curve); $1.7$ (dot-dashed curve). Here $\sigma=0.01$, $\kappa=2.5$\ and
$M=0.75$.}%
\label{fig9}%
\end{center}
\end{figure}

\section{Soliton characteristics}

\label{investigation}

Having explored the existence domains of electron-acoustic solitons, subject
to the constraints of the plasma model, we now turn to consider aspects of the
soliton characteristics. We have numerically solved Eq. (\ref{eq2_37}) for
different representative parameter values, in order to investigate their
effects on the soliton characteristics. We point out that, varying different
parameters, we have found only negative potential solitons, regardless of the
value of $\kappa$ considered. This is not altogether unexpected, as it has
been found in a number of examples that, in contrast to the Cairns
model,~\cite{Cairns1995}, the kappa distribution does not lead to reverse
polarity acoustic solitons.~\cite{Baluku2008,Saini2009}

Figure \ref{fig5} shows the variation of the Sagdeev pseudopotential
$\Psi(\phi)$ with the normalized potential $\phi$, along with the associated
pulse solutions (soliton profiles), for different values of the temperature
ratio, $\sigma=T_{c}/T_{h}$ (keeping $\beta=1.3$, $\kappa=2.5$\ and Mach
number $M=0.75$, all fixed). The Sagdeev potential well becomes deeper and
wider as $\sigma$ is increased. We thus find associated increases in the
soliton amplitude and in profile steepness (see Fig.~\ref{fig5}b). Thermal
effects therefore are seen to amplify significantly the electric potential
disturbance at fixed $M$.

Figure \ref{fig6}a shows the Sagdeev pseudopotential $\Psi(\phi)$ for
different values of $\kappa$. The electrostatic pulse (soliton) solution
depicted in Fig. \ref{fig6}b is obtained via numerical integration. The pulse
amplitude $|\phi_{\mathrm{m}}|$ increases for lower $\kappa$, implying an
amplification of the electric potential disturbance as one departs from the
Maxwellian. Once the electric potential has been obtained numerically, the
cool-electron fluid density (Fig. \ref{fig6}c) and velocity disturbances (Fig.
\ref{fig6}d) are determined algebraically. Both these disturbances are
positive in this case, and again, for lower $\kappa$ values, the profiles
reflecting the compression and the increase in velocity are steeper but narrower.

We recall that as various parameter values are varied, the true acoustic speed
in the plasma configuration, $M_{1}$, also varies. As solitons are inherently
super-acoustic, it is clear that the effect of a changing true acoustic speed
could mask other dependences. Hence it is also desirable to explore soliton
characteristics as a function of the propagation speed $M$, measured relative
to the true acoustic speed, $M_{1}$. This ratio, $M/M_{1}$, thus represents
the ``true" Mach number. It has been shown that for any plasma made up of
barotropic fluids, arbitrary amplitude solitons satisfy $\partial\Psi/
\partial M <0$,~\cite{Verheest2010,Verheest2010a,Verheest2010b} from which it
follows that $\partial\phi_{m} /\partial M >0$, where $\phi_{m}$ is the
soliton amplitude. Thus one expects that the soliton amplitude is an
increasing function of $M/M_{1}$. This is true for both KdV solitons (small
amplitudes, propagating near the sound speed) -- ``taller is faster''-- and,
in principle, also for fully nonlinear (Sagdeev) pulses (where the soliton
characteristics can only be found numerically.~\cite{Saini2009,Baluku2010a})
In Fig. \ref{fig8}, we have plotted the soliton amplitude $|\phi_{\mathrm{m}%
}|$ as a function of the ratio $M/M_{1}$, for a range of values of the
parameter $\kappa$. Clearly, the amplitude increases linearly with $M/M_{1}$
for all values of $\kappa$. The two plots for $\kappa\leq2.5$ both cover the
full range up to $M=M_{2}$. However, although we deduce from earlier figures
that $M_{2}$ increases with $\kappa$, we see that the endpoints of the plots
for $\kappa\geq3$ occur at decreasing values of $M/M_{1}$, and indeed
decreasing maximum amplitudes $\phi_{m}$. That occurs as, for the chosen
values of $\beta$ and $\sigma$, $M_{2}$ exceeds unity for $\kappa\geq3$, and
we have truncated the curves at the point where $M=1$, to remain within the
range defined by the plasma model.

The effect of the hot-to-cool electron density ratio, $\beta$ on the soliton
characteristics
is shown in Fig. \ref{fig9}. We see that the soliton excitations are amplified
and profiles steepened (the Sagdeev potential well becomes wider and deeper),
as the density of the hot (nonthermal) electrons is increased (i.e., for
higher $\beta$), viz., keeping $\kappa$, $\sigma$ and $M$ fixed. Furthermore,
an increase in the number density of the hot electrons also leads to an
increase in the perturbation of both density $n$, and velocity $u$ of the cool
electrons (figure omitted).

\section{Conclusion}

\label{conclusion}

In this article, we have performed a thorough linear and nonlinear analysis,
from first principles, of electron acoustic excitations occurring in a
nonthermal plasma consisting of hot $\kappa$-distributed electrons, adiabatic
cool electrons, and immobile ions.

First, we have derived a linear dispersion relation, and investigated the
dependence of the dispersion characteristics on the plasma environment (degree
of `suprathermality' through the parameter $\kappa$, plasma composition, and
thermal effects).

Then, we have employed the Sagdeev pseudopotential method to investigate large
amplitude localized nonlinear electrostatic structures (solitary waves), and
to determine the region in parameter space where stationary profile solutions
may exist. Only negative potential solitons were found. The existence domain
for solitons was shown to become narrower in the range of solitary wave speed,
with an increase in the excess of suprathermal electrons in the hot electron
distribution (stronger `suprathermality', lower $\kappa$ value). The
dependence of the soliton characteristics on the hot electron number density
(through the parameter $\beta$) and on the hot-to-cool electron temperature
ratio $\sigma$, were also studied. A series of appropriate examples of
pseudopotential curves and soliton profiles were computed numerically, in
order to confirm the predictions arising from the study of existence domains.

It may be added that ionic motion/inertia, here neglected, may also be
included for a more accurate description, but is likely to have only minor
quantitative effects.

We note, for completeness, that very recently two related papers have appeared
with a scope apparently similar to that of the present article, viz., Refs.
\cite{Younsi2010, Sahu2010}. A word of comparison may therefore be appropriate
here, for clarity. The latter authors~\cite{Younsi2010} indicate that they are
using a form of kappa distribution from one of the pioneering papers in the
field.~\cite{Thorne1991} Unfortunately, their expression for the
characteristic speed $\theta$ does not agree with the standard
expression~\cite{Thorne1991}, and thus the hot electron density does not take
the usual form, Eq. (5).~\cite{Baluku2008} Further, they use values of
$\kappa\geq0.6$, i.e., well below the standard $\kappa$-distribution cut-off
of 3/2. Hence their results do not apply to the standard form of $\kappa$
distribution.~\cite{Hellberg2009}

As regards the other paper,~\cite{Sahu2010} it did not consider existence
domains at all (apart from an indirect mention of an upper limit in $M$, as
commented on in Section \ref{existence} above). Further, no account is taken
in the paper of the possible effects of Landau damping on sustainable
nonlinear structures, and a number of the figures relate to values of $\beta$
(called $\alpha$ in the paper) which lie in the unphysical, damped range.

We note that Ref.~\onlinecite{Sahu2010} has also carried out a ``small
amplitude" calculation yielding double layers. However, when evaluated
numerically, these turn out to be well beyond the range of small amplitude,
and that raises some doubts about the validity of the results (their Figures
8-10). Further, let us take together their Figures 2 and 8, and consider the
case of $\kappa=3$ and $\alpha=0.2$ (i.e., our $\beta$). The latter is, of
course, a value for which the linear wave is likely to be strongly Landau
damped. It appears from the figures that a soliton occurs at $M=1.1$ with
amplitude $\sim0.9$, while a double layer occurs for $M=2.0$ with amplitude
$\sim0.7$. This combination of data does not satisfy the analytically-proven
requirement that $\partial\Psi/ \partial M <0$%
,~\cite{Verheest2010,Verheest2010a,Verheest2010b}. There is no obvious reason
why that should be the case, and there thus seems to be an error in at least
one of these two figures.

Finally, we point out that we have not sought double layers in our
calculations. However, we would be surprised if they did occur, as they are
usually found as the upper limit to a sequence of solitons for a polarity for
which there is no other limit. In this case there clearly is an upper limit
for negative solitary waves, arising from the constraint $F_{2}(M)=0$. It is
in principle possible for a double layer to occur at a lower value of $M$, and
be followed by larger amplitude solitons at higher $M$, until the upper fluid
cutoff such as a sonic point or an infinite compression cutoff is
reached.~\cite{Baluku2010} However, such behaviour depends on the the Sagdeev
potential having a fairly complicated shape, with subsidiary local maxima, and
we have not observed these for this model.

We are not aware of any experimental studies with which these theoretical
results may be directly compared. However, it has previously been shown that
wave data may be used to obtain an estimate for $\kappa$, thus acting as a
diagnostic for the distribution function~\cite{Hellberg2000,Vinas2005}.

Similarly, in this case there are a number of indicators amongst our results
which experimenters may wish to consider when interpreting observations. Thus,
for instance, a lower normalized phase velocity of the linear
electron-acoustic wave than would be predicted by a Maxwellian model (see
Fig.~\ref{fig1}) could be used to evaluate $\kappa$.

Secondly, from Fig.~\ref{fig2} one sees that in low-$\kappa$ plasmas the range
of normalized soliton speeds is both narrower and of larger value than one
would expect for a Maxwellian. Thus, if solitons are found with normalized
speeds around $M \simeq0.4$, these can be understood only by allowing for
additional suprathermal electrons (lower $\kappa$). Further, from
Fig.~\ref{fig2} it follows that Maxwellian electrons give rise to a cutoff in
the density ratio $\beta\simeq1$, and hence solitons observed in such plasmas
can only be explained in terms of lower $\kappa$.

From Fig.~\ref{fig6} we note that at fixed values of the normalized
soliton speed, $M$, the amplitudes of the perturbations of the
normalized potential, cool electron density and cool electron speed
due to the solitary waves all increase with decreasing $\kappa$.
This is related to the increase of the true Mach number $M/M_{1}$
for smaller $\kappa$, as the phase velocity $M_{1}$ is decreased.
Thus larger disturbances are likely to be associated with increased
suprathermality.

Finally, turning to Fig.~\ref{fig8}, two effects are observed: At fixed true
Mach number, $M/M_{1}$, the soliton amplitude decreases with decreasing
$\kappa$ (increasing suprathermality). Despite that, the maximum values of
soliton amplitude is found to occur not for a Maxwellian, but for the
relatively low-$\kappa$ values of around 2.5-3. Thus, again, large observed
amplitudes are likely to be associated with a low-$\kappa$ plasma.

Hence, as shown above, these results could assist in the understanding of
solitary waves observed in two-temperature space plasmas, which are often
characterized by a suprathermal electron distribution.

\section*{Acknowledgements}

AD warmly acknowledges support from the Department for Employment and Learning
(DEL) Northern Ireland via a postgraduate scholarship at Queen's University
Belfast. The work of IK and NSS was supported via a Science and Innovation (S
\& I) grant to the Centre for Plasma Physics (Queen's University Belfast) by
the UK Engineering and Physical Sciences Research Council (EPSRC grant No.
EP/D06337X/1). The work of MAH is supported in part by the National Research
Foundation of South Africa (NRF). Any opinion, findings, and conclusions or
recommendations expressed in this material are those of the authors and
therefore the NRF does not accept any liability in regard thereto.

\end{document}